\documentclass[aps,prl,twocolumn,superscriptaddress,byrevtex,amsmath,amssymb]{revtex4}
\usepackage{epsfig}
\begin{document}
\title{Coexistence of ferromagnetism and singlet superconductivity via kinetic exchange}
\author{Mario Cuoco}
\affiliation{Unit\`a I.N.F.M. di Salerno, Dipartimento di Fisica
``E. R. Caianiello'', Universit\`a di Salerno, I-84081 Baronissi
(Salerno), Italy}\affiliation{Centre de Recherches sur les Tr\`es
Basses Temp\'eratures associ\'e \`a l'Universit\'e Joseph Fourier,
C.N.R.S., BP 166, 38042 Grenoble-C\'edex 9, France}

\author{Paola Gentile}
\affiliation{Unit\`a I.N.F.M. di Salerno, Dipartimento di Fisica
``E. R. Caianiello'', Universit\`a di Salerno, I-84081 Baronissi
(Salerno), Italy}

\author{Canio Noce}
\affiliation{Unit\`a I.N.F.M. di Salerno, Dipartimento di Fisica
``E. R. Caianiello'', Universit\`a di Salerno, I-84081 Baronissi
(Salerno), Italy}\affiliation{Unit\`a I.N.F.M. di
Salerno-Coherentia, Salerno, Italy}
\begin{abstract}
We propose a novel mechanism for the coexistence of metallic
ferromagnetism and singlet superconductivity assuming that the
magnetic instability is due to kinetic exchange. Within this
scenario, the unpaired electrons which contribute to the
magnetization have a positive feedback on the gain of the kinetic
energy in the coexisting phase by undressing the effective mass of
the carriers involved into the pairing. The evolution of the
magnetization and pairing amplitude, and the phase diagram are
first analyzed for a generic kinetic exchange model and then are
determined within a specific case with spin dependent bond-charge
occupation.
\end{abstract}
 \maketitle
The problem of the interplay between ferromagnetic (FM) and
superconducting (SC) long range order has been recently attracting
new interest due to the discovery of superconductivity in
ferromagnetic metals, UGe$_{2}$\cite{saxe00},
ZrZn$_{2}$\cite{pfei01}, URhGe\cite{aoki01}, and in
rutheno-cuprate RuSr$_2$RECu$_2$O$_8$ compounds, with RE=Eu or Gd
\cite{tall99}.

The investigation of ferromagnetic superconductors started by
analyzing the case of two interacting subsystems: one formed by
localized spins or aligned magnetic impurities which produces the
ferromagnetic background, the other composed by itinerant
electrons which gives rise to the superconductivity. Within this
framework, early works \cite{abri60,clog62,chan62} focused on
singlet superconductivity in presence of a spin-exchange field,
showing that it can exist only below a critical value of the
magnetic coupling. Hence, with the purpose to increase the
threshold of the critical spin-exchange, it was suggested a finite
momentum pairing state coexisting with the ferromagnetic
order\cite{fuld64,lark64}. In this pairing configuration, commonly
indicated as the Fulde-Ferrell-Larkin-Ovchinnikov state, the
subtle balance between the condensate energy of the Cooper pairs
with a finite center-of-mass momentum and the Zeeman energy,
related to the magnetic moments of depaired itinerant electrons,
was able to account for a superconducting-ferromagnetic (SF) phase
in presence of spin-exchange higher than the zero-momentum pairing
state.

Nevertheless, in the above mentioned materials, a new
phenomenology seems to arise if compared to the conventional case
of a metal with magnetic impurities. Indeed, it has been suggested
that (i) ferromagnetism and superconductivity are cooperative
phenomena, (ii) the FM state is due to itinerant electrons, (iii)
the same electrons participate in both the FM and SC order, and
(iv) for the case of systems with two types of carrier responsible
for the SC and FM phases separately, there are interesting
cooperative effects due to competing charge- and spin-exchange
coupling. Theoretical studies along those directions have been
recently performed by considering the occurrence of ferromagnetism
and superconductivity in either
singlet\cite{blag99,suhl01,abri01,cuoc03} and/or
triplet\cite{kirk01,walk02,mach01} channel of pairing. In
particular, as far as the SC state with singlet {\it s}-wave
pairing is concerned, it has been shown\cite{karc01} that its
coexistence with weak itinerant ferromagnetism can be obtained
within a single band model, where the ferromagnetic order is
driven by the same electrons that participate in the formation of
Cooper pairs. One crucial aspect of such analysis concerns the
stability of the SF state. The loss of condensation energy due to
the depaired electrons which produce a non zero total
magnetization, is not compensated by the energy gain of the
magnetic exchange, thus the SF state turns out always to be
energetically unfavorable against the non-magnetic SC one even if
{\it s}- and {\it d}-wave symmetry or finite-momentum pairing
state are considered\cite{shen03}.

In this letter, we propose a novel mechanism for the coexistence
of superconductivity and ferromagnetism within a single band
model. The new and crucial ingredient is that the metallic
ferromagnetism is not due, as in the previous studies, to a rigid
shift in the positions of the majority and minority spin bands
(i.e. Stoner model), but it is a consequence of a change in the
relative bandwidth of electrons with up and down spin
polarization. In this circumstance, the gain of energy comes from
the undressing of the mass for the majority spins which induces a
bandwidth enlargement and in turn lowers the kinetic
energy\cite{hirs89}. When the pairing interaction is switched on,
the interplay between the gain in kinetic energy, due to the
relative change of the majority and minority spin bands, and the
condensation energy which would tend to pair all the electrons
becomes crucial for the stability of the SF state. It turns out
that such a phase is the most favorable only if a suitable tuning
of the ratio between the density of depaired and paired states is
reached, so to optimize the balance in the kinetic and potential
energy gain. We will analyze this possibility in two steps: I) the
case in which the variation between the mass of up and down spin
electrons is arbitrary modified without referring to any
microscopic mechanism is firstly considered; II) a specific
tight-binding model in which off-diagonal Coulomb interactions are
responsible for an asymmetric bond-charge distribution for each
spin channel, is then investigated. For both cases, the attractive
interaction will be assumed of BCS-type.

{\it Model I}. Let us start from a model Hamiltonian which
contains a local attractive potential and itinerant electrons with
a spin dependent mass:
\begin{eqnarray}
\nonumber
H_{I}&=&-t \sum_{\langle ij,\sigma \rangle} (2 w_{\sigma}) (c^{\dagger}_{i \sigma} c_{j\sigma} + H.c.)-
g \sum_i c^{\dagger}_{i \uparrow} c^{\dagger}_{i \downarrow} c_{i \downarrow} c_{i \uparrow} \\
&&-\mu \sum_{i\sigma}
c^{\dagger}_{i \sigma} c_{i\sigma}
\label{hamI}
\end{eqnarray}
\noindent where $c^{\dagger}_{i \sigma}(c_{i \sigma})$
creates~(destroys) an electron with spin $\sigma$ at the site $i$,
$w_{\sigma}$ is a positive term that controls the renormalization
of the mass for electrons with spin $\sigma$ (the factor $2$ being
introduced for convenience), $\mu$ is the chemical potential, $t$
is the hopping amplitude which defines the bare bandwidth, and $g$
is the pairing coupling being effective only in a shell of
amplitude $2 \omega_c$ around the Fermi surface as in the usual
BCS theory. It is worth pointing out that in this case the mass
dressing and undressing can be generated both via the coupling to
a background of spin, as in double-exchange systems, or to
dynamical processes intrinsically generated by electron
correlations.

The introduction of a pairing amplitude, after the mean-field
decoupling of the attractive term, brings to the following
diagonal expression for the Hamiltonian:
\begin{eqnarray}
H_{I-MF}&=&\sum_{k} \left(E_{k}^{\alpha} \alpha^{\dagger}_{k} \alpha^{}_{k} +
E_{k}^{\beta} \beta^{\dagger}_{k} \beta^{}_{k} \right)+
E_0
\\
E_0&=&\sum_k \left[-E_{k}^{\beta}+(2 w_{\downarrow} \epsilon_{k}-\mu)\right] +
\frac{\Delta^2}{g},
\label{hamBOG}
\end{eqnarray}
\noindent where $\epsilon_{k}=-t \sum_{\delta} \exp{(i k\cdot
\delta)}$ is the bare dispersion with $\delta$ being a vector
connecting a site to its nearest neighbors, and $\Delta=\sum_k g
\langle c^{}_{k \uparrow} c^{}_{-k \downarrow} \rangle$ is the
pairing amplitude, respectively. The field operators $\alpha_{k}$,
$\beta^{}_{k}$ correspond to fermionic excitations with
quasiparticle dispersion
\begin{eqnarray}
E_{k}^{\alpha,\beta}&=& \pm a~\epsilon_{k}+\sqrt{(b~\epsilon_{k}-\mu)^2+\Delta^2 }.
\label{dispersion}
\end{eqnarray}
\noindent $b=w_{\uparrow}+w_{\downarrow}$ and
$a=w_{\uparrow}-w_{\downarrow}$ being the average and half the
difference of the spin mass renormalization. It is immediately
apparent that the kinetic exchange amplitude is proportional to
$a$ while the value of $b$ determines the order of magnitude of
the average kinetic energy. Still, the sign of the total
magnetization ($M$) follows that of $a$, while $b$ is always
positive. Thus, without any loss of generality one can focus only
on the case with $a \geq 0$, the other one being just derived by
inverting the direction of $M$.

The equations for the
pairing amplitude and the magnetization have the following form
\begin{eqnarray}
\Delta&=&\frac{g~\Delta }{2} \sum_k \frac{1-n_k^\alpha-n_k^\beta}
{\sqrt{(b~\epsilon_{k}-\mu)^2+\Delta^2 }}\\
M&=&\frac{\mu_B}{2} \sum_k \left(n_k^\beta-n_k^\alpha\right),
\label{eqGM}
\end{eqnarray}
\noindent where $n_k^{\alpha,\beta}$ are the Fermi distributions
in the momentum space of the fermionic fields having
$E_{k}^{\alpha,\beta}$ as energy dispersion relations,
respectively, and $\mu_B$ is the Bohr magneton. Let us now discuss
under what conditions the system can accommodate a coexistence of
SC and FM order. The analysis will be restricted to the zero
temperature limit and furthermore we will assume that the density
of electrons $n$ is self-consistently fixed by a positive value of
the chemical potential $\mu$. The results can be symmetrically
extended within the same procedure to the negative $\mu$ case.
Hence, considering that $b,a,\Delta,\mu ~\geq 0$ and $b\geq a$, it
is possible to show that $n_k^{\alpha}=0$ as $E_{k}^{\alpha}$ is
always positive, while $n_k^{\beta}=1$ when $\lambda_{k,-}\leq
\epsilon_{k} \leq \lambda_{k,+}$, where $\lambda_{k,\pm}=\frac{b
\mu \pm \sqrt{ a^2 (\Delta^2+\mu^2)-b^2 \Delta^2 }}{b^2-a^2}$. The
$\lambda_{k,\pm}$ define a range on the positive side
$\epsilon_k$, which is different from zero only when $a$ and $b$
are linked to the amplitude of $\Delta$ and $\mu$ in a way to
fulfill the relation $\frac{a}{b}>
\frac{\Delta}{\sqrt{\Delta^2+\mu^2}}$. If the previous inequality
holds, single particle states form within the gap and contribute
to give a finite magnetization. Then, depending on their density
the system can allow for a state with coexisting SC and FM order.
In the limit of $\mu\rightarrow 0$ ($n\sim 1$) or $a \rightarrow
0$ ($M \sim 0$), there are no real solutions for $\lambda_{k,\pm}$
so that one ends up with the usual BCS-like state with zero
magnetization. In the other cases, the self-consistent equations
for the gap amplitude and the number of electrons have to be
solved to determine the conditions for the existence of SF state.
For such purpose, it has been used a model density of states given
by $N(\epsilon)=1/2w$ (with $w=2t$) if $-w \leq \epsilon \leq w$.
We notice that, contrary to the case of the Stoner model, a
non-trivial ferromagnetic solution can be got without the need of
a curvature in the density of states close to the Fermi
level\cite{hirs89}. The analysis of the solutions is performed by
fixing the value of $b$ and $n$ and studying how the pairing
amplitude, the energy and the magnetization are modified by
changes of $a$. Due to the formation of unpaired electrons within
the gap, the value of the effective average mass in the SF state
($b_{sf}$) is modified with respect to the case of zero
magnetization ($b_{sc}$). Hence, there might occur two distinct
cases depending on the microscopic mechanism which controls the
relative change of the majority and minority spin bandwidth: 1)
undressing of the average effective mass, i.e. $\delta b>0$; 2)
dressing of the average effective mass, i.e. $\delta b<0$, $\delta
b\equiv(b_{sf}-b_{sc})/{b_{sc}}$ being the relative percentage
variation.

\begin{figure}[t]
\centerline{\psfig{figure=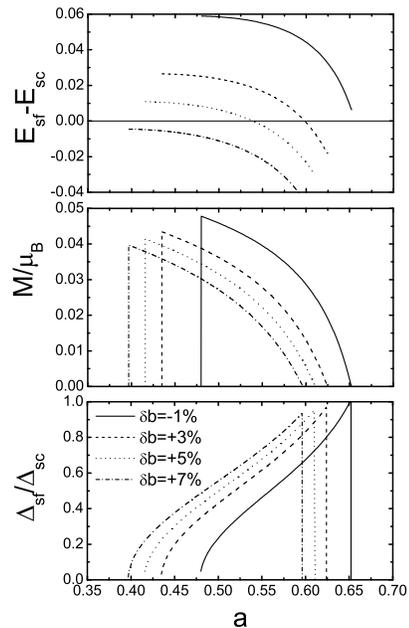 ,width=6.cm}}
  \caption{From the bottom to the top: superconducting gap, total
  magnetization, and relative energy between the SF and the SC phase,
  respectively, as a function of $a$,
  for different values of the average effective
  mass renormalization. The density of electrons has been fixed to $n=1.2$ and the coupling strength is $g/t=1$.}
  \label{fig1}
\end{figure}

In Fig.\ref{fig1} is shown the comparison between the energy of
the SF and the SC state, together with the behavior of the
magnetization and the pairing amplitude as a function of the
relative shift between the mass of up and down spin electrons. For
simplicity, we have fixed $b_{sc}=1$ and we have studied how the
difference of energy is modified in the cases (1) and (2). Changes
in $b_{sc}$ produce only quantitative but not qualitative
differences in the results. As shown in the top panel of
Fig.\ref{fig1}, only when $\delta b$ is positive and for a
suitable density of the depaired states, controlled by the
amplitude of $a$, the system can be stabilized in a SF phase. When
the average mass increases, though the effective amplitude pairing
grows and in turn lowers the condensation energy, the loss in the
kinetic energy of the electrons that participate in the pairing
cannot be counterbalanced to get a stable SF state. It is worth
pointing out that the transition from the SF state to the SC one
is of first order for the magnetization but it is continuous for
the pairing amplitude by moving from small to large values of $a$,
while it is reversed in the opposite $a$ direction. Nevertheless,
depending on the amplitude of the undressing $\delta b$, the
transition from the SF state can be first order type in both the
order parameters. Finally, due to the peculiar link between
$\Delta$ and $\mu$, we notice that the change in the density of
electrons modifies the region where the SF solution exists. In
particular, the interval of solutions shrinks if one moves towards
the half-filling case, whereas a large band undressing is required
to stabilize the SF phase.

{\it Model II}. Now we consider an explicit tight-binding model
where the bandwidth change depends on the bond-charge occupation
in each spin channel\cite{hirs89}. The model Hamiltonian has the
following form:
\begin{eqnarray}
\nonumber && H_{II} = -t \sum_{<ij>\sigma}\left(c^{\dag}_{i
\sigma}c_{j
\sigma}+ H.c \right)+ U \sum_{i}n_{i \uparrow}n_{i \downarrow}\\
\nonumber &&+V\sum_{<ij>}n_{i}n_{j} +J\sum_{<ij>\sigma
\sigma^{\prime}} (c^{\dag}_{i \sigma}c^{\dag}_{j
\sigma^{\prime}}c_{i \sigma^{\prime}}c_{j \sigma}+H.c)
\\ \nonumber && + J^{\prime}\sum_{<ij>} (c^{\dag}_{i \uparrow}c^{\dag}_{i
\downarrow}c_{j \downarrow}c_{j \uparrow}+ H.c)- g \sum_{i} (
c^{\dag}_{i\uparrow}c^{\dag}_{i\downarrow}c_{i\downarrow}c_{i\uparrow}+H.c)\\&&-\mu
\sum_{i\sigma} c^{\dagger}_{i \sigma} c_{i\sigma} \label{hamII}
\end{eqnarray}
\noindent where $U$ and $V$ represent the on-site and
nearest-neighbor Coulomb repulsion, respectively, and the
parameters $J$ and $J^{\prime}$ describe nearest-neighbor exchange
and pair hopping processes, while $g$ is the BCS pairing. By
applying the Hartree-Fock decoupling, one can obtain the
expression of the bare quasi-particle dispersion in terms of a
bond charge quantity $I_{\sigma}=\langle c^{\dagger}_{i\sigma}
c_{j\sigma} \rangle$ that in the momentum space is given by
$I_{\sigma}=\sum_k n_{k\sigma} \left(\frac{-\epsilon_k}{w}\right)$
in a way that the kinetic part of the Hamiltonian
(\ref{hamII})(neglecting the Stoner exchange) reads
$H_{II-MF}=\sum_{k\sigma} \left[(1-2 j_{1} I_{\sigma}-2 j_{2}
I_{-\sigma}) \epsilon_{k}-\mu\right] n_{k \sigma}$. \noindent It
is possible to show that the values of $j_1$ and $j_2$ are linked
to the $V$, $J$, and $J^{\prime}$ interactions by means of the
following relations: $j_1=(J-V)/{w}$ and
$j_2=(J+J^{\prime})/{w}$\cite{hirs89}. Hence, by following the
same procedure as in {\it Model I}, it is convenient to introduce
the parameters $a$ and $b$, which now has to be determined
self-consistently via the following relations: $b=
\left[1-(j_1+j_2)(I_{\uparrow}+I_{\downarrow})\right]$ , $
a=(j_2-j_1)(I_{\uparrow}-I_{\downarrow})$

\noindent Still, $I_{\sigma}$ depends on $a$ and $b$ by means of
the occupation number of the $\alpha$ $(\beta)$ bands, via the
following relations:
\begin{eqnarray*}
n_{k\uparrow}&=&v^2_k+u^2_k (n^{\alpha}_k+n^{\beta}_k)-n^{\beta}_k\\
n_{k\downarrow}&=&v^2_k+u^2_k
(n^{\alpha}_k+n^{\beta}_k)-n^{\alpha}_k .
\end{eqnarray*}
\noindent where $u^2_k=(\frac{1}{2})(1+\frac{(b
\epsilon_k-\mu)}{\sqrt{(b \epsilon_k-\mu)^2+\Delta^2}})$ and
$v^2_k=1-u^2_k$ are the coefficients of the Bogoliubov
transformation.

Few comments are worth mentioning on the possible solutions and on
the effect of depaired electrons in the amplitude of the average
mass $b$ within the SF state. First of all, the amplitude of $a$
and $b$ is determined by $j_{-}=(j_2-j_1)$ and $j_{+}=(j_1+j_2)$,
respectively. Moreover, the sign of $j_{+}$ controls whether the
starting value in SC state for $b$ is larger or smaller than 1,
while $j_{-}$ sets the sign of the magnetization. As far as the
stability is concerned, like in the {\it Model I}, it is crucial
to see whether the introduction of depaired electrons in the SF
phase reduces or increases the average effective mass with respect
to that of the SC state. Hereafter, we will assume that $j_{\pm}$
are positive, which is consistent, in the weak coupling regime,
with realistic values of the microscopic couplings~($J,J^{'},V$)
above introduced.

Let us now analyze the dependence of $b$ in the case of the SF and
SC phase. One can show that $b_{sf}$ gets undressed in the
coexisting phase, thus its value is larger than $b_{sc}$ when
$n^{\beta}_k \neq 0$. This result can be deduced by writing down
the explicit expression of the ground state energy of the SF and
SC phases. The expectation value of the bond charge term on the
unpaired electrons gives a negative contribution to
$I_{\uparrow}+I_{\downarrow}$ which is not present in the
non-magnetic SC case, thus increasing the average kinetic energy
and consequently the value of $E_0$ in the coexisting state.

\begin{figure}[t]
\centerline{
\psfig{figure=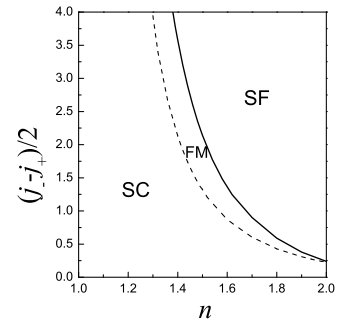,width=5cm}
}
  \caption{Phase diagram relative to the {\it Model II}, showing the
  transition(full line) from the SF to the SC state, and that one
  from a paramagnetic
  to FM state when $\Delta=0$ (dashed line), as
  the difference $(j_{-}-j_{+})$ and the density of electrons are varied. Due to the
  particle-hole symmetry, the
  part for $n$ in the range $\left[0,1\right]$ is just symmetrically related.}
  \label{fig2}
\end{figure}

In Fig. \ref{fig2} it is reported the phase diagram for
$j_{+}=0.5$, $g/t=0.25$ obtained by varying $j_{-}$ with respect
to the density. Of course, by modifying the amplitude of $j_{+}$
one can span the phase diagram for all the possible values of
$j_1$ and $j_2$. We have checked that a change in $j_{+}$ yields a
shift in the critical line but does not give any qualitative
change in the phase diagram. The phenomenology of the transition
between the SC and SF state is the same of that observed for the
{\it Model I}. Indeed, when one goes through the critical line,
the magnetization has a jump to its possible maximal value which
depends on $\Delta$ and $\mu$, while the SC gap grows continuously
from zero. For completeness, in Fig.\ref{fig2} it is also reported
the line of transition from a paramagnetic to a FM state in the
case of absence of superconductivity. The shape of the region of
stability in the $[(j_{-}-j_{+}),n]$ diagram can be understood by
noticing that the kinetic exchange is given by
$k_{ex}=\epsilon_{\uparrow}-\epsilon_{\downarrow}=2~a~\epsilon$
and $a=2 j_{-} m (n-1)$ (for the pure ferromagnetic case). Thus
approaching the half-filling limit ($n \rightarrow 1$) one has
that $a\rightarrow 0$ and consequently the kinetic exchange goes
to zero so that it is needed an infinite value of $j_{-}$ to get a
spin polarized state. This consideration explains the asymptotic
behaviour of the critical line as we get close to the half-filling
limit where it becomes more and more difficult to have a
magnetized state because the difference in the bond charge
occupation for opposite spin becomes close to zero thus requiring
high values of the coupling constants to create a charge
unbalance.

In conclusion, we have studied the occurrence of ferromagnetism
and {\it s}-wave singlet superconductivity within a single band
model where the magnetic moments are due to a kinetic exchange
mechanism, both for a generic and for a specific case with
bond-charge coupling. It has been shown that the depaired
electrons play a crucial role in the energy balance and that only
when their dynamical effect is such to undress the effective mass
of the carriers which participate in  the pairing, then the SF
phase can be stabilized. As far as the specific case of a
bond-charge kinetic mechanism is concerned, we have shown that the
phase diagram has a peculiar dependence on the density of
carriers, and that only when the system is far from the limit of
exact particle-hole symmetry, the SF phase can be obtained without
going to very large couplings.

M.C. acknowledges support from the European program "Improving
Human Potential". The authors thank Maria Teresa Mercaldo for
helpful comments and valuable discussions.

\end{document}